\documentclass[english]{article}
\usepackage[T1]{fontenc}
\usepackage[latin9]{inputenc}
\usepackage{graphicx}

\makeatletter
\newcommand{\lyxaddress}[1]{
\par {\raggedright #1
\vspace{1.4em}
\noindent\par}
}

\makeatother

\usepackage{babel}
\begin{document}

\title{The questionable impact of population-wide public testing in reducing
SARS-CoV-2 infection prevalence in the Slovak Republic}

\author{Jozef \v{C}ern\'{a}k\footnote{jozefcernak@gmail.com}}
\maketitle

\lyxaddress{Department of Nuclear and Sub-Nuclear Physics, Faculty of Science,
Institute of Physics, Pavol Jozef Šafárik University in Košice, Košice,
Slovak Republic}
\begin{abstract}
Mina and Andersen, authors of the Perspectives in Science: \textquotedbl{}COVID-19
Testing: One Size Does Not Fit All\textquotedbl{} have referred to
results and adopted conclusions from recently published governmental
report Pavelka \textit{et al.} \textquotedblleft The effectiveness
of population wide, rapid antigen test based screening in reducing
SARS-CoV-2 infection prevalence in Slovakia'' without critical consideration,
and rigorous verification. We demonstrate that the authors refer to
conclusions that are not supported by experimental data. Further,
there is a lack of objective, independent information and studies
regarding the widespread, public testing program currently in force
in the Slovak Republic. We offer an alternative explanation of observed
data as they have been provided by the Slovak Republic government
to fill this information gap. We also provide explanations and conclusions
that more accurately describe viral spread dynamics. Drawing from
available public data and our simple but rigorous analysis, we show
that it is not possible to make clear conclusions about any positive
impact of the public testing program in the Slovak Republic. In particular,
it is not possible to conclude that this testing program forces the
curve down for the SARS-CoV-2 virus outbreak. We think that Pavelka
\textit{et al.} did not consider many fundamental phenomena in their
proposed computer simulations and data analysis - in particular: the
complexity of SARS-CoV-2 virus spread. In complex spatio-temporal
dynamical systems, small spatio-temporal fluctuations can dramatically
change the dynamics of virus spreading on large scales. 
\end{abstract}
INTRODUCTION:

Mina and Andresen in the paper \cite{Mina} refer to mathematical
models that incorporate relevant variation in viral loads and test
accuracy \cite{Larremore}. On that basis, they suggest that - with
frequent, large-scale sampling of a population - detection of herd
effects was possible. The authors \cite{Mina} referred to the public
testing currently in effect in the Slovak Republic \cite{Pavelka}.
Unfortunately, they have adopted research conclusions that are not
supported by appropriate mathematical models, which require additional
variables and factors. Further, the authors of the initial report
\cite{Pavelka} have not rigorously compared the results with actual
measurement data from the Slovak Republic over a longer term window,
i.e. a few weeks before and after public -wide testing (31. October
- 1. November 2020). Our independent and simple analysis of WHO available
global data from Slovak Republic (S1) shows quit the opposite effect
and throws into doubt the conclusions the authors \cite{Pavelka}. 

\begin{figure}
\includegraphics[width=10cm]{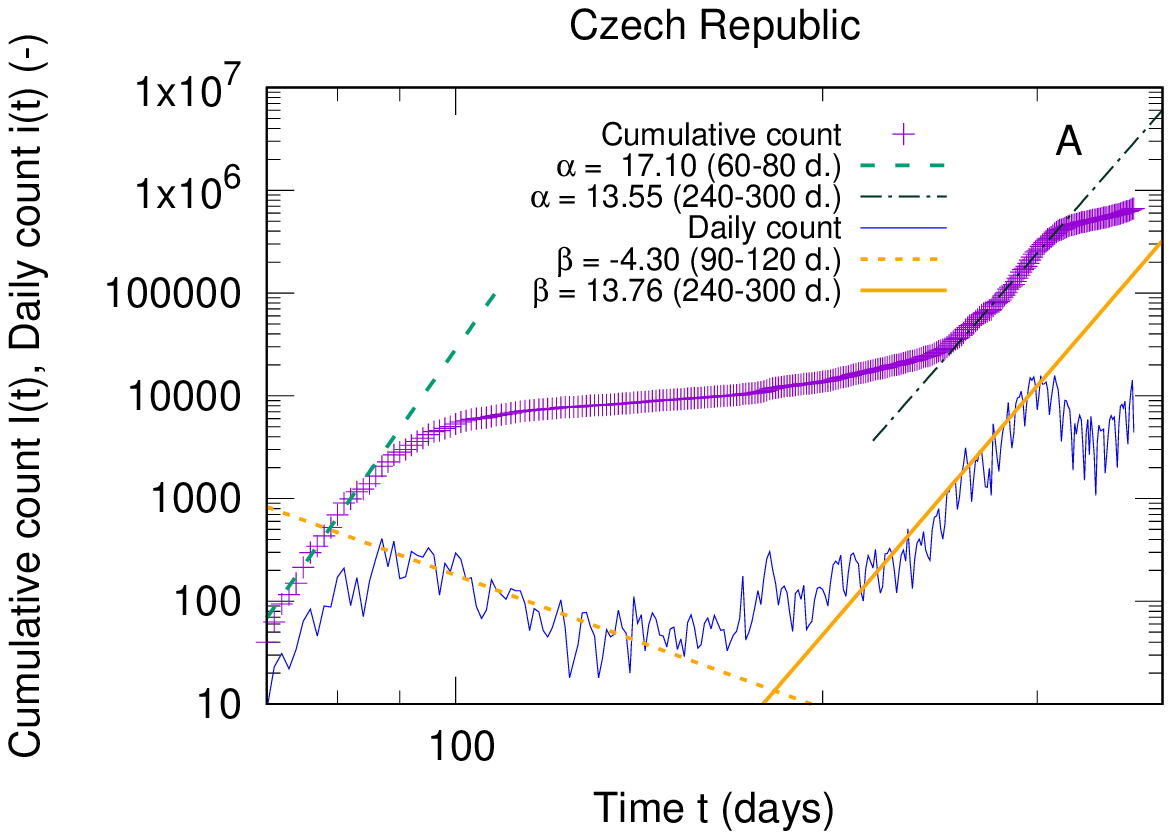}

\includegraphics[width=10cm]{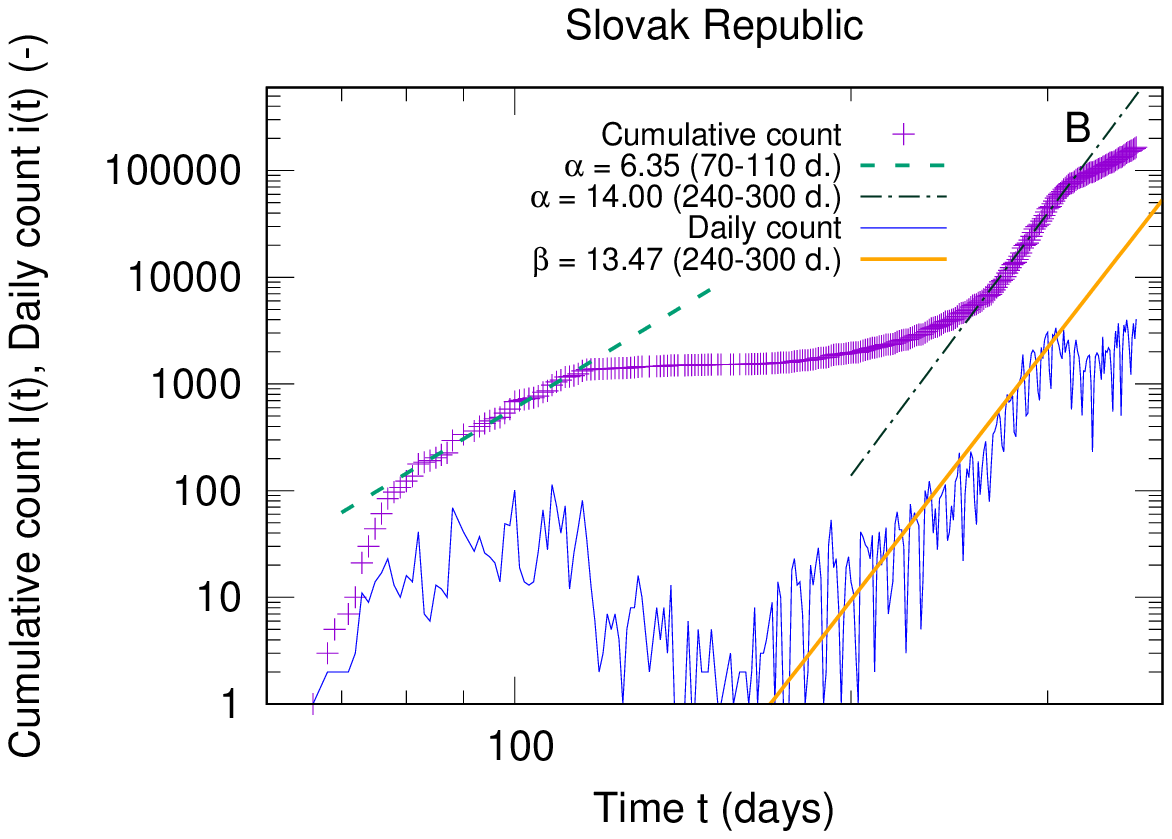}

\caption{\label{fig:Scaling}Scaling properties of cumulative count $I(t)\sim t^{\alpha}$
and daily count of infected cases $i(t)\sim t^{\pm\beta}$ in (\textbf{A})
Czech Republic and in (\textbf{B}) Slovak Republic, time $t$ is measured
from 3. January 2020 (S1). }
\end{figure}

RESULTS:

We have analyzed the scaling properties \cite{Tizzoni} of daily count
$i(t)$ as well as cumulative count $I(t)$ of infected cases where
$t$ is time in days during the first and second SARS-CoV-2 virus
waves. Our results Figure \ref{fig:Scaling} show scaling properties
of $i(t)\sim t^{\pm\beta}$, $I(t)\sim t^{\alpha}$ where $\alpha$
and $\beta$ are scaling exponents. Double logarithmic scales Figure
\ref{fig:Scaling} are much more suitable to demonstrate scaling properties
and to identify significant changes of virus spread dynamics, for
example to recognize outbreak waves as well as the dynamics of outbreak
growth and decay during a time of the wave. 

A power law decay of daily count of infected cases $i(t)\sim t^{-\beta}$
shows that a decay of outbreak follows a slow dynamics and can take
a long time depending on both an exponent $\beta$ and a number of
daily infected cases $N$ in tipping point of daily count of infected
cases $i(t)$ (in a preparation to publish). 

\begin{figure}
\includegraphics[width=10cm]{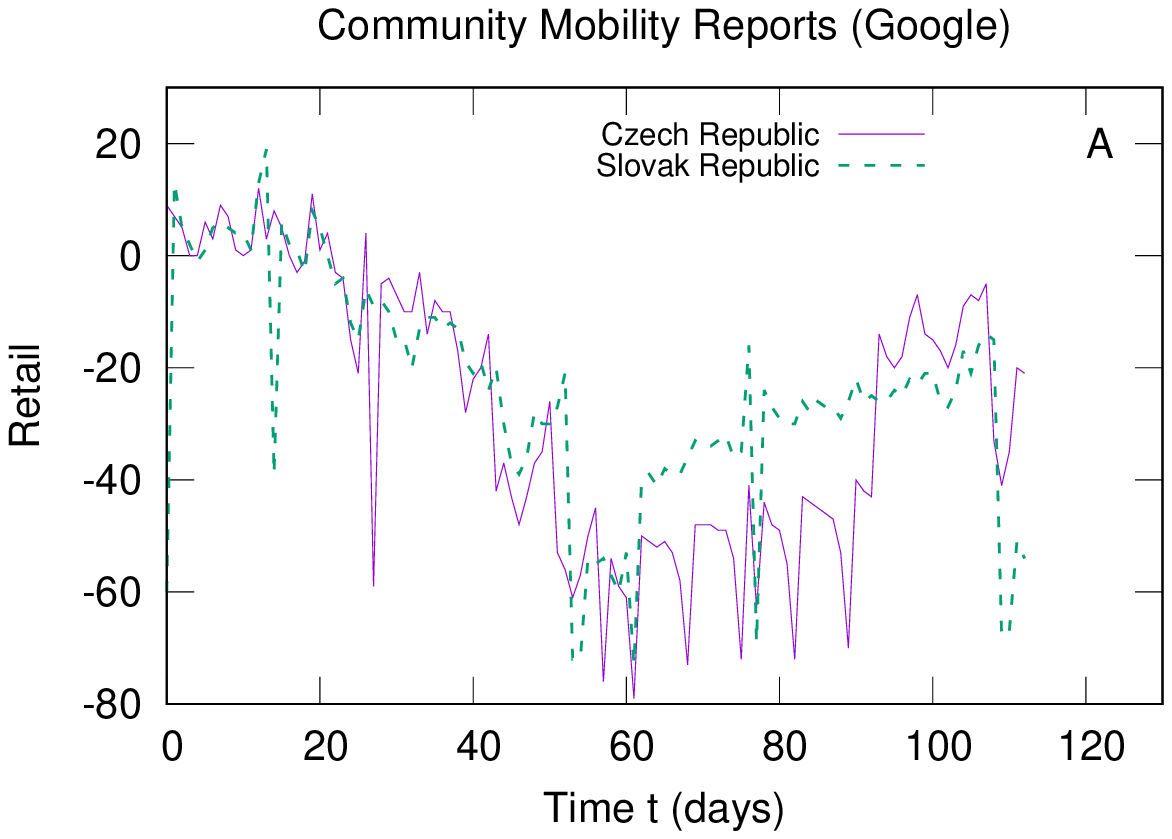}

\includegraphics[width=10cm]{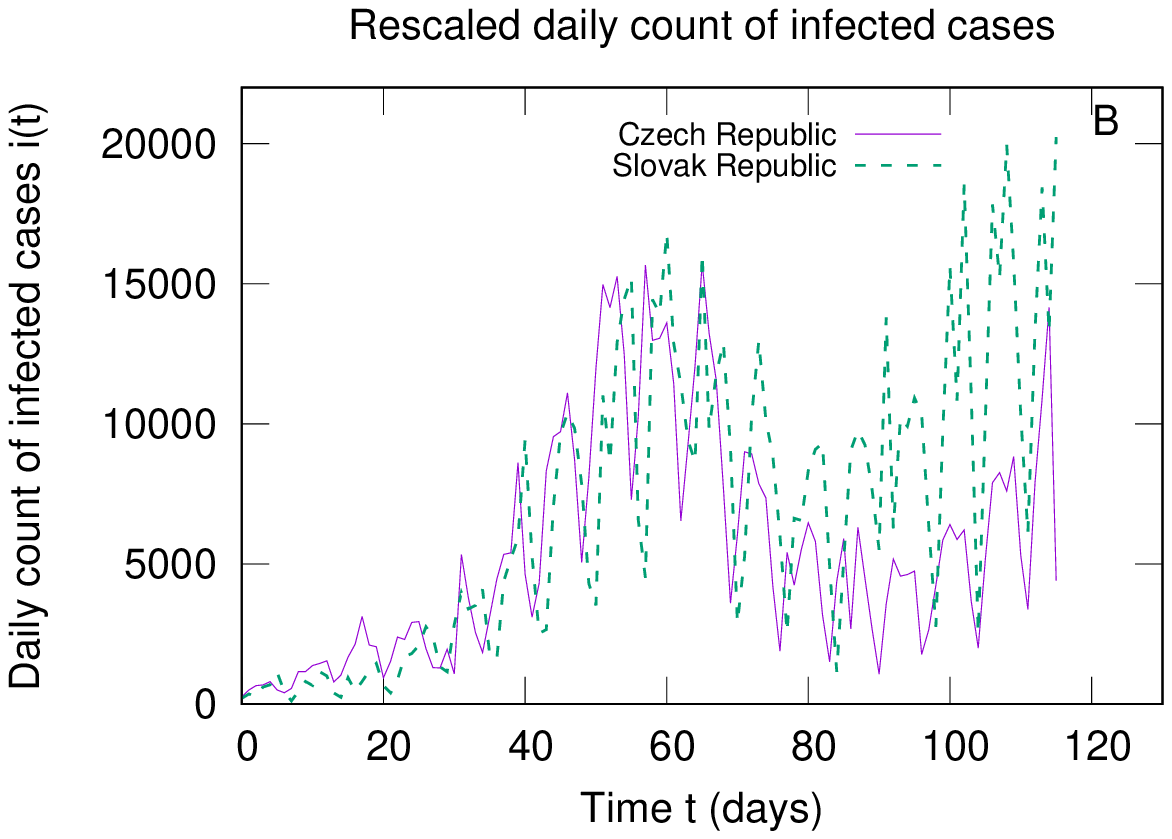}

\caption{\label{fig:tight}(\textbf{A}) Community Mobility Reports in the Czech
Republic and Slovak Republic provided by Google (S2). (\textbf{B})
Linear-linear plot of rescaled daily count of infected cases $i(t)$
in the Czech Republic and Slovak Republic. Time $t$ is mesaused from
1. September 2020 (S2). }
\end{figure}

We have analyzed only one component of mobility Figure \ref{fig:tight}
(S2): retail and recreation, that carries important information about
the effectiveness of public policy measures i.e. demonstrating that
these measures decrease average mobility and therefore the average
number of daily personal contacts. 

In Figure \ref{fig:Scaling} we can identify in these neighbor countries
a common tipping point of daily count of infected cases $i(t)$ on
1. November 2020. We compare a temporal evolution of a retail mobility
Figure \ref{fig:tight} (\textbf{A}) and rescaled daily count of infected
cases Figure \ref{fig:tight} (\textbf{B}) in Czech Republic and Slovak
Republic). You can see common features of retail mobility and daily
count of infected cases before the tipping point and quit different
features of mobility as well as daily count of infected cases subsequent
the tipping point. Retail mobility Figure \ref{fig:tight} (\textbf{A})
and daily count of infected cases Figure \ref{fig:tight} (\textbf{B})
clearly demonstrate that, if countries applied similar public policy
measures to decrease mobility, that the dynamic of virus spread has
similarly decayed in both countries. After public-wide testing in
the Slovak Republic (31. October-1. November 2020), mobility dynamics
Figure \ref{fig:tight} (\textbf{A}) as well as rescaled daily count
of infected cases Figure \ref{fig:tight} (\textbf{B)} dramatically
changed in the Slovak Republic. The rescaled daily count of infected
cases in Slovak Republic shows a much more higher daily count of infected
cases as when both countries shared similar public policy measures
to control low mobility. 

DISCUSSION:

Our criticism is focused on the work of Pavelka \textit{et al.} \cite{Pavelka}.
We think that Pavelka \textit{et al.} \cite{Pavelka} made several
conceptual mistakes during their analysis of available data and their
assumptions regarding governmental public policy measures. Most egregiously,
they did not discuss the potential effects of false negative results
in that report \cite{Pavelka}. SD Biosensor claims that a combined
negative agreement with PCR tests is $99.7\%$ (see the company data
sheet regarding tests results in Switzerland). Based on data provided
by the authors \cite{Pavelka} and SD Biosensors, we estimate $15600$
false negative tests (i.e. the infected cases that were falsely evaluated
as negative cases). Shortly after the public testing phase, the Slovak
Republic government permitted the free movement of tested persons
Figure \ref{fig:tight} (\textbf{A}), while it has drastically restricted
the free movement of healthy persons that opted to not participate
in the public testing program. 

This increase in mobility of tested population (Figure \ref{fig:tight}
(\textbf{A})) - including persons who have false negative test results
- would logically suggest an uncontrolled increase of infection in
all regions of the Slovak Republic within the following 7- 14 days
after testing. This has now been confirmed by publicly available data
Figure \ref{fig:Scaling} (\textbf{B}) and Figure \ref{fig:tight}
(\textbf{B}). We note that it is necessary to consider the long incubation
period of SARS-COV-2 virus and the average time when first syndromes
could occur \cite{Hu}. The authors \cite{Pavelka} have not discussed
the important impacts of other measures that were applied before public
testing began, for example a decrease in mobility - very similar to
that experienced in Czech Republic and Slovak Republic Figure \ref{fig:tight}
(\textbf{A}) who share the same tipping point of daily count of infected
cases at 1. November 2020 Figures \ref{fig:Scaling} (\textbf{A}),
(\textbf{B}) and \ref{fig:tight} (\textbf{B}). We note that, at this
tipping point, the reproductive number has been $R<1$ and public
testing program in the Slovak Republic had been started. Importantly,
the author's computer simulations \cite{Pavelka} did not take into
account the influx of new infected cases from abroad due to periodic
- and massive - migration of work forces between the Slovak Republic,
Czech Republic and other countries. 

The history of the SARS-CoV-2 outbreak in the Czech Republic and Slovak
Republic - prior to the Slovak's Republic testing program Figures
\ref{fig:Scaling} (\textbf{A}), (\textbf{B}), and \ref{fig:tight}
(\textbf{B})\textbf{ -} show that these countries were strongly coupled,
with similar daily counts of infections and similarly decreasing infection
trends due to low-mobility and other important public-policy measures
taken in the Slovak Republic and neighboring countries. Subsequent
to the \textquotedbl{}tipping point\textquotedbl{}, the decreasing
trend in daily infections in the Slovak Republic virtually stopped
within a few days. Daily count of infected cases Figure \ref{fig:tight}
(\textbf{B}) started again to increase. This is in contrast to the
situation in the Czech Republic Figures \ref{fig:Scaling} (\textbf{A}),
(\textbf{B}) and \ref{fig:tight} (\textbf{B}). This directly demonstrates
that the public testing program has not had any positive effect on
daily infection rates. We show in Figure \ref{fig:tight} (\textbf{B}),
that public testing - in an environment where tests are not precise
and there is a relative high mobility of tested persons (many with
false negative test results) - can initiate new outbreaks. Our interpretation
of available data is entirely contrary to the interpretations and
conclusions as presented in \cite{Pavelka} and uncritically adopted
by other authors \cite{Mina}. We are confident that our conclusions
are well supported by other authors who have investigated the SARS
outbreak and mathematically investigated the impacts of quarantine
and other public-policy measures in the past \cite{Lip}. These authors
concluded that quarantine appears to have formed the most effective
basis for control in several countries and should be equally effective
on a smaller scale, likely contributing to the prevention of major
outbreaks in other countries. On the other hand, in the absence of
such effective measures, SARS has the potential to spread very widely.
Considerable effort will be necessary to implement such measures in
those settings where transmission is ongoing, but such efforts are
essential to quell local outbreaks and reduce the risk of further
global dissemination \cite{Lip}. We think that in the context of
large scale populations, it is very difficult to control the effectiveness
of wide public quarantine (personal remark: i.e. without drastic violation
of human rights) due to the complexity of virus spread as well as
of personal contact interactions \cite{Tizzoni}.

CONCLUSIONS:

We believe that a detailed and correct analysis of SARS-CoV-2 virus
spread in the Czech and Slovak Republics is very important and could
be useful for a better understanding of dynamic of SARS-CoV-2 outbreak.
Both the Czech Republic and Slovak Republic have been successful in
stopping the first wave of SARS-CoV-2 outbreak Figure \ref{fig:Scaling}.
On the other hand, the countries have not be able to smoothly manage
the second wave. The current approaches to manage the outbreak in
these countries are quite different. In the case of the Czech Republic,
the main tools are to limit mobility and increase testing, while the
Slovak Republic engages in a model of very intensive and frequent
testing virtually everywhere \cite{Larremore,Pavelka} with a relative
high mobility allowed in tested populations. Since the tipping point
(1. November 2020), the data does not support any positive impact
of this approach in the Slovak Republic Figure \ref{fig:tight} (\textbf{B}).
We think that this is due to the complexity of virus spread, rapid
and uncoordinated shifts in public policy, non-optimal communication
with citizens and a very low effectiveness of quarantine control on
large scales \cite{Lip}

\section*{Acknowledgment}

We thank Geir Helgesen for valuable discussion and Ben Dowling for
reading the manuscript.

\section*{Supplementary materials}

\end{document}